\documentclass[10pt]{iopart}

\usepackage{epsf}
\usepackage{graphics}
\usepackage{graphicx}

\begin{document}

\title{In-plane transport anisotropy in BSCCO-Ag multi-filamentary tapes}

\author{A Borroto$^{1}$, A S Garc{\'i}a-Gordillo$^{1}$, L Del R{\'i}o$^{1,2}$, M Arronte$^{3,4}$ and E Altshuler$^{1}$}

\address{$^1$ Superconductivity Laboratory, Physics Faculty-IMRE, University of Havana, 10400 Havana, Cuba}
\address{$^2$ Physics Department, McGill University, Montreal, Quebec, Canada H3A 2T8}
\address{$^3$ BRALAX, S. de RL., Tampico, TAMPS, M{\'e}xico}
\address{$^4$ Technological Laser Laboratory, IMRE, University of Havana, 10400 Havana, Cuba}
\ead{ealtshuler@fisica.uh.cu}
\vspace{10pt}
\begin{indented}
\item[]April 2015
\end{indented}

\begin{abstract}
Composite structures such as High-$T_c$ multi-filamentary tapes display a complex anisotropy arising from the combination of the ``intrinsic" anisotropy of the Bi-2223 grains, and that associated to the superconducting phase distribution in the superconductor-metal composite, as well as cracks and other defects. In this paper we characterize the ``in-plane" anisotropy of $\textrm{BSCCO-Ag}$ tapes, i.e., the difference between the transport properties along the longitudinal axis and those along the transverse direction also lying on the wide face of the tape. In particular, we demonstrate that the dissipation associated to transport along the transverse direction approaches that of the longitudinal direction as the temperature or the current increase, which may be relevant to transport applications in situations where the superconducting properties have significantly degraded.
\end{abstract}

%
\vspace{2pc}

\submitto{\SUST}
%
%
\ioptwocol

\section{Introduction}

Anisotropy is a well known ``intrinsic" feature of several high $T_c$ superconductors that emerges from their crystalline structure. For example, the critical current along the a-b planes of the structure can be much higher than that along the c-direction \cite{Martin1989,Nomura1990}. That anisotropy is observed in a straightforward fashion in single crystals or epitaxial films.

\begin{figure}[b]
    \vspace{0.2cm}
     \begin{center}
     \includegraphics[height=2.5in, width=2.8in]{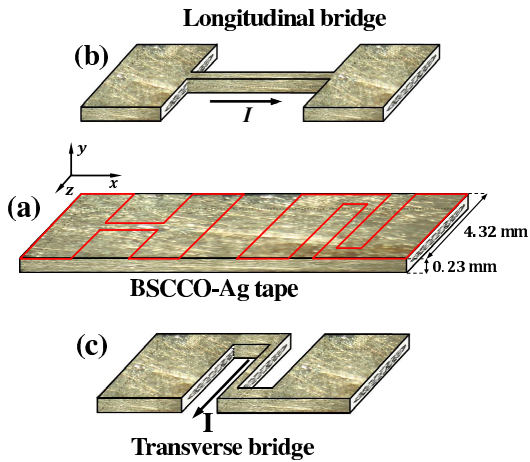}
     \end{center}
     \vspace{-0.7cm}
    \caption{Sketches of longitudinal and transverse bridges. (a) Whole BSCCO-Ag tape, the bridges were cut from it. (b) Longitudinal bridge. (c) Transverse bridge.}
    \label{fig:circuit}
\end{figure}

However, superconducting samples are typically inhomogenous at a morphological level much coarser than the crystals they are made of: bulk superconductors \cite{Clem1988,Altshuler1991,Altshuler1999,Batista-Leyva2000,Batista-Leyva2003,Treimer2012}, wires \cite{Corato2009}, tapes \cite{Ogawa2003,Zhang2012}, and coated superconductors \cite{Trommler2012,Colangelo2012,Solovyov2013} display different levels of ``granularity". For example, BSCCO-Ag tapes consist in filaments made from superconducting grains that are very well connected along the wide face of the tape, embedded in a non-superconducting matrix. For those, there is a large ``in-plane" anisotropy between the longitudinal and transverse directions of the tape (x and z axis in figure \ref{fig:circuit}) \cite{Borroto2013,Borroto2014} --in addition to the better known anisotropy between the properties along the x-z plane, and those along the y-axis (figure \ref{fig:circuit}). The anisotropy between the longitudinal and transverse directions of a multi-filamentary superconducting composite is relevant to the AC losses \cite{Wilson2008} and influences the availability of ``lateral escape ways" to the current in the presence of transverse cracks \cite{Cai1998,Akiyama2014}, which has been rarely reported in the literature \cite{Borroto2013,Koblischka1997,Bobyl2000,Bobyl2002}.

In this paper, we study in detail the ``in-plane" anisotropy of BSCCO-Ag tapes. First, we demonstrate that the transition to the dissipative regime is wider along the transverse direction that along the longitudinal one. In addition, using an anisotropy parameter defined ad-hoc, we show that the dissipation along the longitudinal direction of the tape approaches that along the transverse direction (both lying on the wide face of the tape) as the temperature and the applied current increase. This transition from high to low in-plane anisotropy is potentially relevant to understand the behavior of superconducting tapes in the high dissipation regime, where transverse cracks and other defects may play an important role.

\section{Experimental details}

We use experimental raw data published in a
previous paper by our group \cite{Borroto2013} as well as some extra
data obtained under similar experimental conditions, briefly
described here. Samples were prepared from a silver-sheathed 61-filament $Bi_2Sr_2Ca_2Cu_3O_{10+x}$ tape
(BSCCO-Ag), measuring $4.32$~mm in width and $0.23$~mm in thickness.
The overall critical (engineering) current of the tape was $I_c = 65$~A at liquid
nitrogen temperature. Narrow bridges with $0.30$~mm of width were
cut from the tape using a laser technique which has been shown not to affect the superconducting properties of the tape
\cite{Borroto2013,Sanchez2007}. Current-voltage curves were measured using four probes.
The current contacts were attached outside the narrow bridges and
far from the voltage contacts, i.e., on the wider areas shown in figures
\ref{fig:circuit}(b) and \ref{fig:circuit}(c), to guarantee that possible current contact
heating had minimal effect on the voltage measurements.

\section{Results and discussion}

\subsection{Critical current and n-index}

Two important parameters characterizing the quality of superconducting tapes are the critical current $I_{c}$ and the $n$-index associated to the $I-V$ curves. It is well known \cite{Bruzzone2004,Ghosh2004,Ochiai2009,Pitel2013} that the transition to the dissipative regime of multifilamentary superconductors can be described by a power law with exponent $n$ given by:
\begin{equation}
 \langle{E}\rangle = \langle{E}\rangle_{c} \left(\frac{I}{I_{c}}\right)^n  \label{powerlaw}
\end{equation}
where $I_{c}$ is the critical current defined by a given electric field criterion $\langle{E}\rangle_{c}$. The criterion of electric field used here was $\langle{E}\rangle_{c}=1 $ $ \mu $V cm$ ^{-1} $. Though some authors use a smaller threshold to determine $I_{c}$ (for example  $\langle{E}\rangle_{c}=0.1 $ $ \mu $V cm$ ^{-1} $), our choice has shown to be suitable for $\textrm{BSCCO}$ superconducting tapes and in small dimensions samples \cite{Ghosh2004,Pitel2013}.

\begin{figure}[t]
    \begin{center}
    \includegraphics[height=2.4in, width=3.2in]{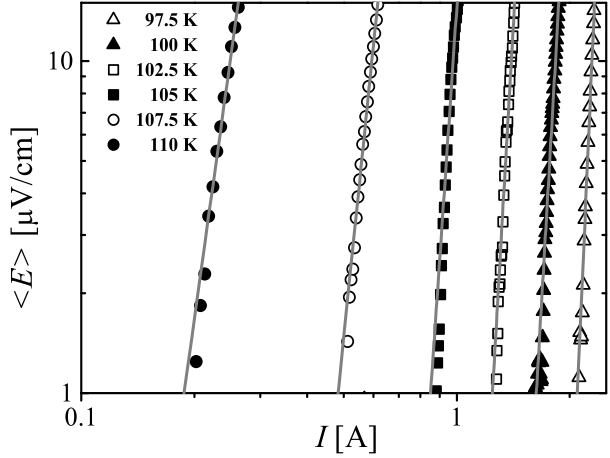}
    \end{center}
    \vspace{-0.6cm}
    \caption{$\langle{E}\rangle-I$ curves in the longitudinal direction obtained with a separation of 2.92 mm between voltage contacts. The gray lines follow (\ref{powerlaw}).
    }
    \label{fig:Elong}
\end{figure}

\begin{figure}[t]
    \begin{center}
    \includegraphics[height=4.8in, width=3.1in]{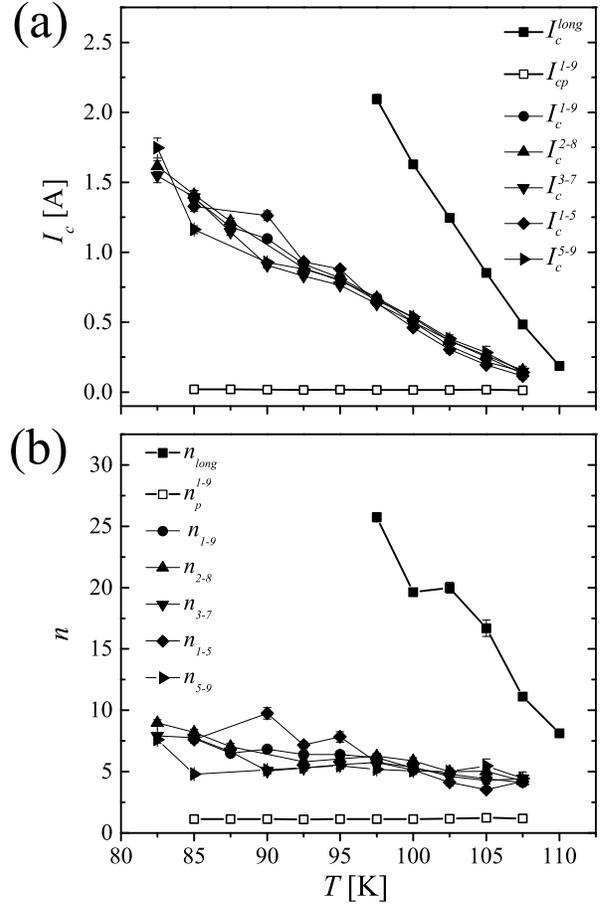}
    \end{center}
    \vspace{-0.6cm}
    \caption{Dependency with temperature of: (a) the critical current and (b) the $n$-index. The dependency with temperature of $I_{c}$ and $n$ in the longitudinal direction is shown with black squares. The empty squares shows the values of $I_{cp}$ and $n_{p}$ obtained through the fitting of (\ref{powerlawcomp}) to the ``linear" region of the $\langle{E}\rangle-I$ curves in the transverse direction. The other curves correspond to the values of $I_{c}$ and $n$ in the transverse direction to different positions of the voltage contacts.}
    \label{fig:Ic-ntrans}
\end{figure}

In figure \ref{fig:Elong} it is shown the $\log \langle{E}\rangle-\log I$ dependence for a longitudinal bridge (see figure \ref{fig:circuit}(b)) for different values of temperature and a separation of $2.92$~mm between voltage contacts. The linear behavior of the curves can be described by (\ref{powerlaw}). If we fit (\ref{powerlaw}) to the $\langle{E}\rangle-I$ curves, using $I_{c}$ and $n$ as fitting parameters, we obtain their dependency with temperature which are shown in figure \ref{fig:Ic-ntrans} as black squares.

\begin{figure}[t]
    \begin{center}
    \includegraphics[height=2.4in, width=3.2in]{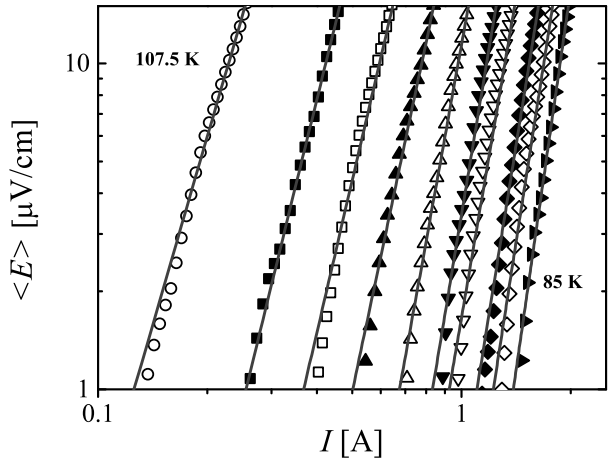}
    \end{center}
    \vspace{-0.6cm}
    \caption{$\langle{E}\rangle-I$ curves in the transverse direction after
subtracting the lineal behavior to the experimental data. The results were obtained with a separation of 2.61 mm between voltage contacts. The curves, from right to left, were obtained from $85$ to $107.5$~K, in steps of $2.5$~K. The gray lines follow (\ref{powerlaw}).}
    \label{fig:Etran}
\end{figure}
In the transverse direction we cannot obtain directly the values of $I_{c}$ and $n$ from the experimental data, as we did in the longitudinal one. This is due to the fact that in the transverse direction there is always dissipation owing to the presence of inter-filament silver in the way of the current \cite{Borroto2013}. As a consequence, (\ref{powerlaw}) is not directly applicable to the experimental data because, even at low applied currents, the $\langle{E}\rangle-I$ curves have a slope, in contrast with the longitudinal direction, where for currents smaller than $I_{c}$, $\langle{E}\rangle$ is negligible \cite{Borroto2013,Borroto2014}. Furthermore, throughout the temperature range studied, there is an interval of currents where the slope of the $\langle{E}\rangle-I$ curves is almost constant. The procedure we followed was to fit to the experimental data, in the region where the $\langle{E}\rangle-I$ dependence is nearly linear, the equation:
\begin{equation}
  \langle{E}\rangle = \langle{E}\rangle_{c} \left(\frac{I}{I_{cp}}\right)^ {n_{p}}   \label{powerlawcomp}
\end{equation}
where $I_{cp}$ is the value of the current where the electric field criterion, $\langle{E}\rangle_{c}=1$ $\mu$V cm$^{-1}$, is reached. So, $I_{cp}$ is the critical current in the transverse direction of the superconductor-silver composite, which is very small due to the fact that in this direction there is always dissipation. After fitting equation (\ref{powerlawcomp}) to the ``linear region" of the $\langle{E}\rangle-I$ curves, it was subtracted from the experimental data.

Figure \ref{fig:Etran} shows the experimental data after subtracting the linear region. These curves exhibit essentially the same behavior of the ones in the longitudinal direction, i.e. for all temperature values there is an interval of currents where the slope of the $\langle{E}\rangle-I$ curves is null (not shown in the graph) and a region where the dependency $\log \langle{E}\rangle-\log I$ is linear. Therefore, it is safe to assume that, by subtracting the linear behavior to the $\langle{E}\rangle-I$ curves in the transverse direction, the extra dissipation has been eliminated.

\begin{figure}[t]
    \begin{center}
    \includegraphics[height=2.9in, width=3.3in]{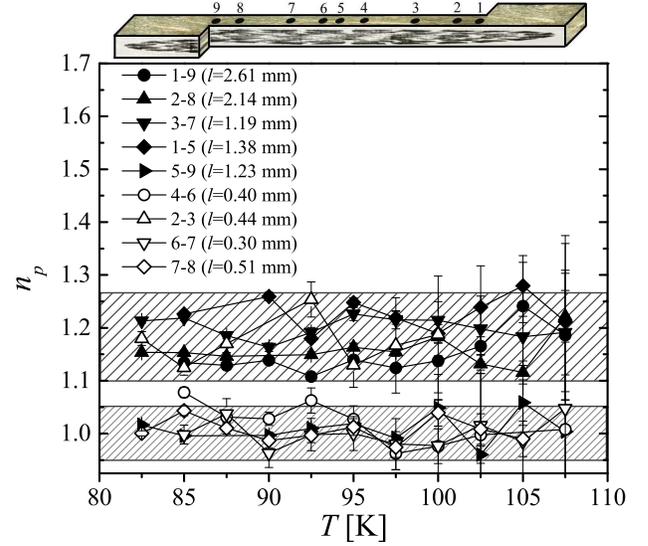}
    \end{center}
    \vspace{-0.6cm}
    \caption{Dependence of $n_{p}$ with temperature corresponding to different positions of voltage contacts along the transverse bridge. On top there is a diagram showing the positions where contacts were put.}
    \label{fig:npindex}
\end{figure}

The same procedure was applied to $\langle{E}\rangle-I$ curves obtained for voltage contacts at different positions along the bridge. Figure \ref{fig:npindex} shows the dependence of the $n_{p}$ coefficient with temperature obtained through the fitting of (\ref{powerlawcomp}) to different positions of voltage contacts along the transverse bridge (on top of figure \ref{fig:npindex} the positions where contacts were put are shown). Notice that if the $\langle{E}\rangle-I$ dependence is really linear, the $n_{p}$ coefficient is equal to $1$. In figure \ref{fig:npindex} we can see that $n_{p}\approx1$ for some contact positions but in other positions $n_{p}>1.1$. The fact that the $n_{p}$ coefficient is greater than $1$ can be associated to the current percolation through the filaments, which we assume responsible of the loss of linear behavior in the $\langle{E}\rangle-I$ curves. This effect is more pronounced for the largest separations between voltage contacts, where current has the chance of describing more complex trajectories.

The $I_{cp}-T$ dependencies were also obtained. In figure \ref{fig:Ic-ntrans}(a) it is shown with empty squares, when the voltage contacts are at the ends of the bridge. Only this configuration is shown to make the figure clearer because for the other contact distances the results are practically the same. We can see, as we said before, that the values of $I_{cp}$ are almost zero.

After subtracting the extra dissipation component to the $\langle{E}\rangle-I$ curves, we fit (\ref{powerlaw}) to the resulting curves. The fittings were done in the same manner as in the longitudinal direction, using an electric field criterion of $\langle{E}\rangle_{c}=1 $ $ \mu $V cm$ ^{-1} $. This way, the $I_{c}-T$ and $n-T$ dependencies were obtained in the transverse direction. These results are shown in figure \ref{fig:Ic-ntrans}. Notice that the values of $I_{c}$ and $n$ obtained for separations between contacts less than $1$~mm were not shown. This is due to the fact that the noise in the electric field rises when the voltage contacts get closer, which makes impossible to fit (\ref{powerlaw}) in the interval of electric fields we are studying. Figure \ref{fig:Ic-ntrans} also shows, with black squares, the dependencies of $I_{c}$ and $n$ obtained in the longitudinal direction. In the transverse direction, both $I_{c}$ and $n$ decrease by increasing temperature as in the longitudinal direction but their magnitudes are smaller.

Due to the fact that the a-b planes of the filaments lie on the wide face of the tape, one might expect small anisotropy in the transport properties between the longitudinal and transverse directions. Nevertheless, in figure \ref{fig:Ic-ntrans}(a) we can see that the critical currents along both directions are different. Notice, however, that the magnitude that must be similar is the critical current density $J_{c}$, because $I_{c}$ depends on the superconducting area perpendicular to the direction in which current flows. This area, even though it can be considered as constant in the longitudinal direction, is highly variable in the transverse one; that is why we are not able to obtain $J_{c}$ in this direction using only the experimental data. However, by means of a model already developed by us \cite{Borroto2014} we have estimated $J_{c}$, which turns out to be very similar in both directions.

Figure \ref{fig:Ic-ntrans}(b) shows that the $n$-index is smaller in the transverse direction than in the longitudinal one. The $n$-index is inversely related to the width of the $V-–I$ transition. So, even though the critical current densities are similar in both directions, the way in which the filaments pass to the mixed state is different in the two directions. This can be understood if we assume that the superconductor is made up of a series of ``regions" which critical currents are not necessarily similar. Then, the potential drop experimentally measured is the result of the sum of the potential drops of each one of these regions, thus determining the shape of the $\langle{E}\rangle-I$ curves. So, it is reasonable to assume that a smaller dispersion of the critical current values gives as a result ``steeper" $\langle{E}\rangle-I$ curves, whose corresponds to a high value of the $n$-index. On the other hand, $\langle{E}\rangle-I$ curves are less ``steep" if there is a greater dispersion of the critical current values, which corresponds to a smaller $n$-index. Therefore, taking into account the experimental results shown in figure \ref{fig:Ic-ntrans}(b), it can be concluded that critical currents along the longitudinal direction are more homogeneous than along the transverse direction.

\subsection{In-plane transport anisotropy}

As we said before, when one speaks about anisotropy in the field of high temperature
superconductors, it is typically associated with the anisotropy
between the a-b planes and the c axis. However, we are dealing here with
the ``in -plane" anisotropy between the longitudinal and transverse
directions relative to the main axis of the tape, which is more
connected with the morphology of the superconductor-silver composite
than with the ``intrinsic" anisotropy of the superconducting filaments
themselves (whose a-b planes lie parallel to the wide face of the
tapes). This anisotropy is of interest in the presence of cracks orthogonal to the direction of the filaments, which interfere with longitudinal current flow over large areas \cite{Cai1998,Akiyama2014}. Based on the $\langle{E}\rangle-I$ curves, we can quantify it as:
\begin{equation}
  A_{lt}=1-\frac{\langle{E}\rangle_{long}}{\langle{E}\rangle_{tran}} \label{Anisotropy}
\end{equation}
where $\langle{E}\rangle_{long}$ and $\langle{E}\rangle_{tran}$ are
the electric fields for the longitudinal and transverse directions
respectively. Then, if there is ``perfect superconductivity" in the
longitudinal direction, $A_{lt}=1$. On the other hand, when the
dissipation in the longitudinal and transverse directions are
equal, $A_{lt}=0$.

\begin{figure}[t]
    \begin{center}
    \includegraphics[height=5.3in, width=3.2in]{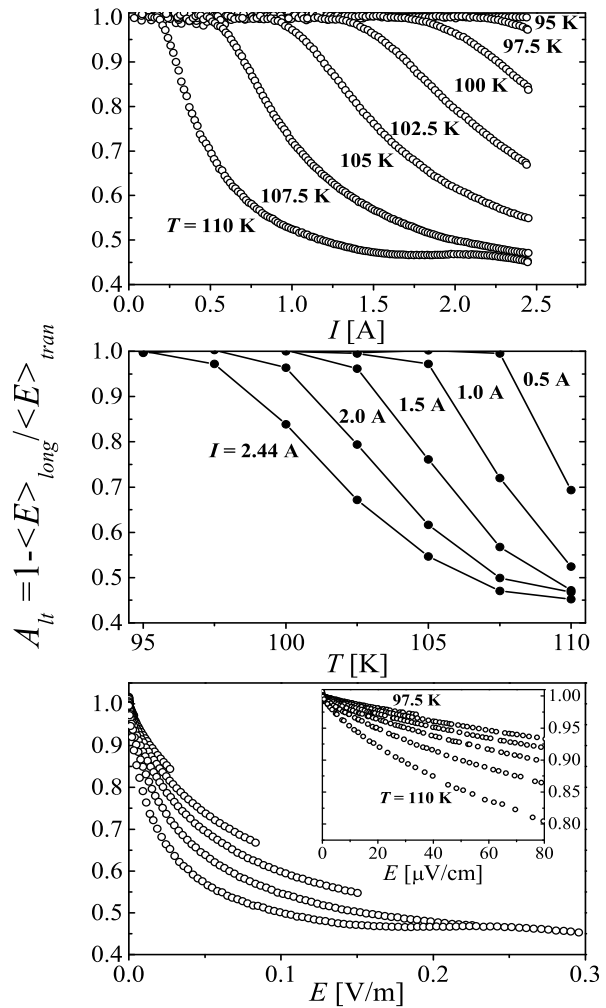}
    \end{center}
    \vspace{-0.6cm}
    \caption{ Anisotropy between longitudinal and transverse directions. Upper panel: anisotropy vs. current for different temperatures. Intermediate panel: anisotropy vs. temperature for different currents. Bottom panel: anisotropy vs. electric field for different temperatures.
    }
    \label{fig:Anis}
\end{figure}

The upper panel in figure \ref{fig:Anis} shows the experimental dependence of
$A_{lt}$  as a function of the applied current for different
temperatures (open circles). In the intermediate and bottom panels of figure \ref{fig:Anis},
the dependence of $A_{lt}$ with the applied current and the electric field in the longitudinal direction are shown. Notice that, as $A_{lt}$ decreases, the
dissipation along the main direction of the tape increases relative
to the dissipation of the transverse one. Following our data, by increasing the temperature, the applied current or the electric field, we see that the contribution of the transverse currents eventually is as relevant as the contribution of the longitudinal ones.
All in all, our results show that the degree of anisotropy at
different temperatures and currents is given by the specific geometry and morphology of the
composite, and can be ultimately understood by means of percolation
simulations, which are in progress.

\begin{figure}[t]
    \begin{center}
    \includegraphics[height=4.in, width=2.8in]{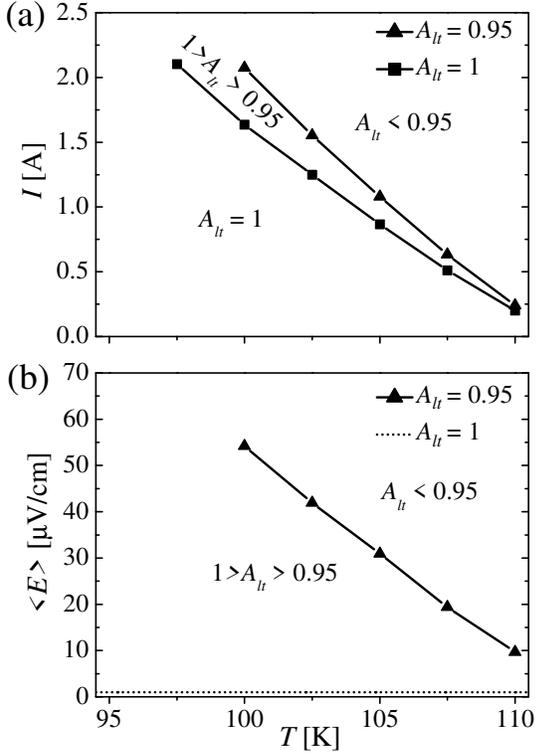}
    \end{center}
    \vspace{-0.6cm}
    \caption{ ``Phase diagrams". (a) Current-temperature plane. The squares show the dependence of the critical current with temperature in the longitudinal direction and the triangles show the dependence of the current with temperature for an $A_{lt}$ value of $0.95$. (b) Electric field-temperature plane. The triangles show the dependence of the electric field with temperature for $A_{lt}=0.95$. The dotted line shows the value $\langle{E}\rangle_{c}=1 $ $ \mu $V cm$ ^{-1} $. In both figures, in the region where $A_{lt}>0.95$ the longitudinal direction is much less dissipative than the transverse one.
    }
    \label{fig:PhasDiag}
\end{figure}

A useful tool to characterize the tape are the phase diagrams shown in figure \ref{fig:PhasDiag}. The diagram shown in figure \ref{fig:PhasDiag}(a) was constructed with the help of the upper panel of figure \ref{fig:Anis}. To do this, an $A_{lt}$ criterion of 0.95 was selected and the dependence of the current with temperature for this criterion was obtained (black triangles). This means that the triangles split the region with $A_{lt}>0.95$ from the region with $A_{lt}<0.95$ in the current-temperature plane (notice that $A_{lt}=0.95$ is equivalent to $\langle{E}\rangle_{tran}=20\langle{E}\rangle_{long}$). Thus, for current and temperature values that belong to the region in which $A_{lt}>0.95$, the longitudinal direction is much less dissipative than the transverse one (the electric field is, at least, $20$ times bigger in the transverse direction than in the longitudinal direction). Nevertheless, in the region where $A_{lt}<0.95$, dissipation in both directions starts to be comparable.

The squares in figure \ref{fig:PhasDiag}(a) show the dependence of the critical current with temperature in the longitudinal direction. Thus, under the curve delimited by the squares, $A_{lt}=1$, because the dissipation in the longitudinal direction is zero, while the transverse direction is always dissipative. Notice that there is a region bounded by square and triangle symbols in which dissipation in the transverse direction is still much higher than dissipation in the longitudinal one.

Figure \ref{fig:PhasDiag}(b) shows a diagram similar to the one already shown in figure \ref{fig:PhasDiag}(a). In this case was used the bottom panel of figure \ref{fig:Anis} and using a criterion of $0.95$ the dependence of the electric field with temperature for this criterion was obtained. This dependence is shown with black triangles. Therefore, below the region delimited by the triangles, the longitudinal direction is much less dissipative than the transverse one. The dotted line in figure \ref{fig:PhasDiag}(b) correspond to $\langle{E}\rangle_{c}=1$ $\mu$V cm$^{-1}$. So, when temperature decreases, the tape is able to work in a wider range of electric fields without ``transverse escapes" of the transport current.
\begin{figure}[t]
    \begin{center}
    \includegraphics[height=2.3in, width=3.2in]{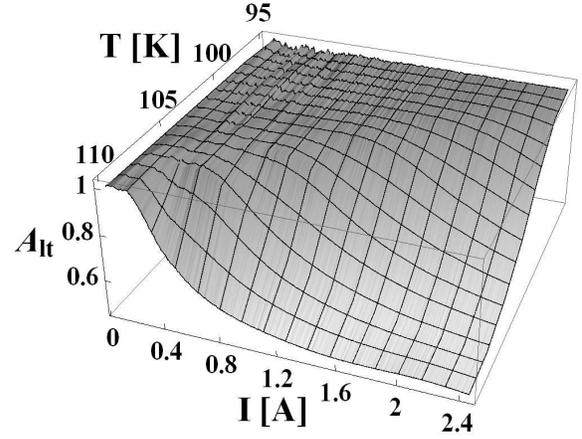}
    \end{center}
    \vspace{-0.6cm}
    \caption{ Temperature and current in-plane anisotropy. As temperature and current increase, $A_{lt}$ decrease, which means that the transverse direction is able to stand a higher current relative to the longitudinal one.
    }
    \label{fig:Alt3D}
\end{figure}

\section{Conclusions}

The critical current and $n$-index in the longitudinal and transverse directions of $\textrm{BSCCO-Ag}$ multifilamentary tapes were investigated. The transition to the dissipative regime is wider in the transverse direction than in the longitudinal one, which is associated to a more inhomogeneous morphology along the former one.

We have quantified the anisotropy of the tapes between their longitudinal and transverse directions through the $A_{lt}$ coefficient --a parameter of relevance if transverse cracks force the current to detour from the main direction of the tape. This coefficient shows a non trivial relation with the temperature, the applied current and the electric field which we represent in the diagram displayed in figure \ref{fig:Alt3D}: as current and temperature increase, $A_{lt}$ decrease, which means that transport currents are more likely to circulate in the transverse direction.

\ack
We thank J. Carrillo and G. S{\'a}nchez-Colina for experimental
help, Z. Han for providing the tapes and R. Packard for support.
E. A. thanks M. \'{A}lvarez-Ponte for inspiration.

\end{document}